\documentstyle[prl,aps,floats,epsfig]{revtex}

\begin{document}
\draft
\renewcommand{\topfraction}{0.8}
\twocolumn[\hsize\textwidth\columnwidth\hsize\csname 
@twocolumnfalse\endcsname
 
\title{Comment on ``Absence of abelian Higgs hair for extremal black holes''}
\author{Filipe Bonjour and Ruth Gregory}
\address{Centre for Particle Theory, 
Durham University, South Road, Durham, DH1 3LE, U.K.
}
\date{4 September 1998}
\maketitle
%
%
\pacs{PACS numbers: 04.70.-s, 11.27.+d, 98.80.Cq \hfill
hep-th/9809029 \hfill DTP/98/63}

\vskip2pc]


The issue of what hair a black hole can wear has been an
ongoing and interesting story. The hair in question 
in this comment is abelian-Higgs hair, which may occur 
because a U(1) vortex can pierce, or even end, on a black hole
\cite{AGK}. 
In the letter of Chamblin et.\ al.\ \cite{CCES} this issue was explored for
charged black holes, and it was argued that while the results of 
\cite{AGK} were qualitatively the same for nonextremal black holes,
in the extremal limit a completely new phenomenon occurred, and
the flux of the vortex was expelled from the black hole, rather like
flux is expelled from a superconductor.
The evidence presented was a set of plots
in which the flux lines of the vortex consistently wrapped
the black hole, even for very thin vortices.
We have re-examined this system, and while we do find flux expulsion for
thick strings, we do not agree that thin vortices are expelled.

The equations for the vortex were 
solved numerically in \cite{CCES}, using a 
technique first described in \cite{AGK}. The basic idea is to discretize
on a grid in $r$ and $\theta$, and to
use a relaxation procedure. The only subtlety is that one of the boundaries --
the horizon -- does not have fixed boundary conditions, but must also
be relaxed to a solution, therefore  there is a choice of
{\it initial} conditions, and it turns out that this is a key issue.
We use exactly the same numerical techniques as
\cite{CCES,AGK}, but with a variety of initial conditions
on the horizon. We chose three representative  data sets corresponding
to setting core, exterior, and an intermediate sinusoidal form for
the vortex fields.

\begin{figure}[htbp]
  \begin{center}
    \epsfig{file=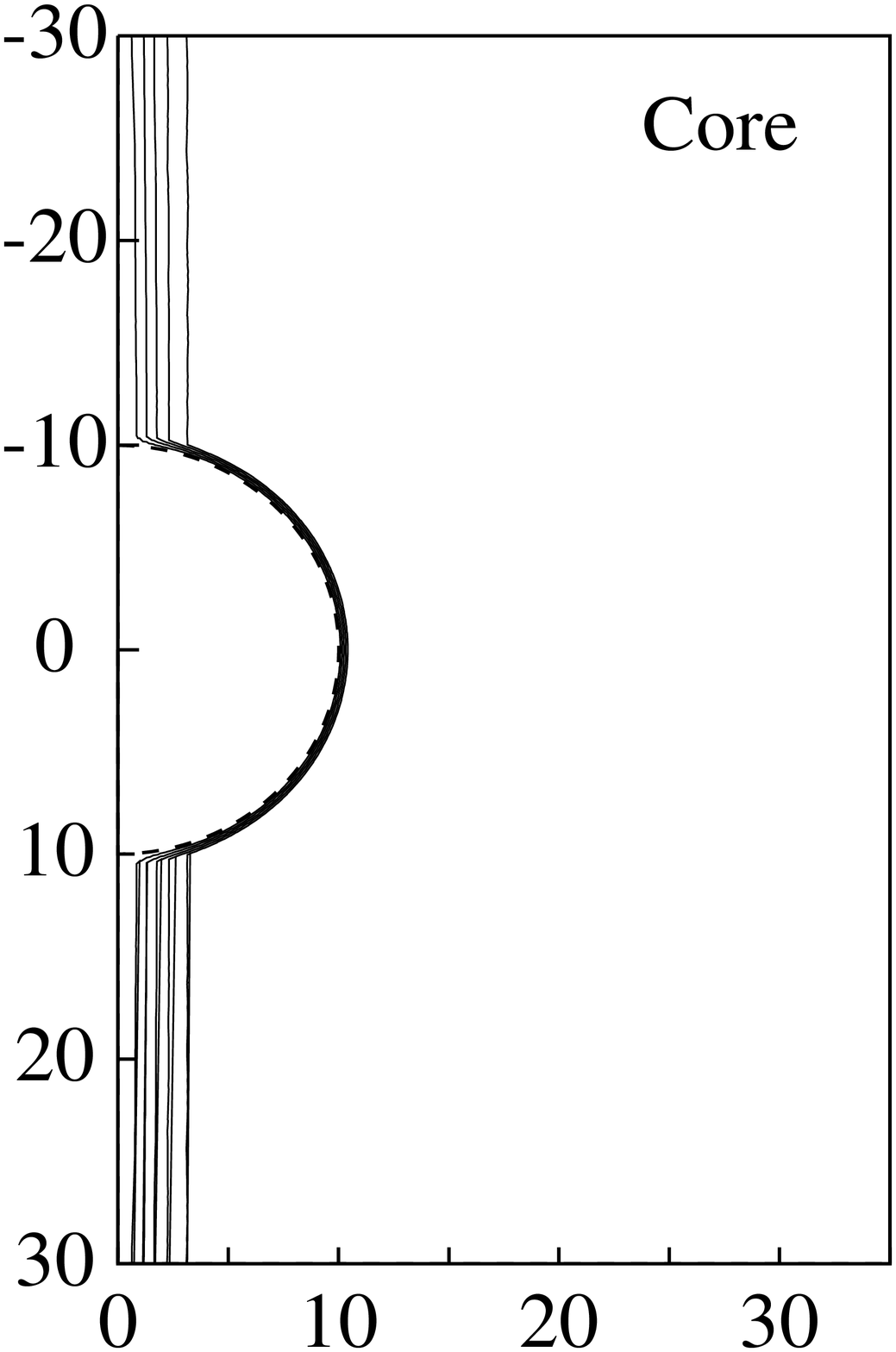,width=3.8cm,angle=180}~%
    \epsfig{file=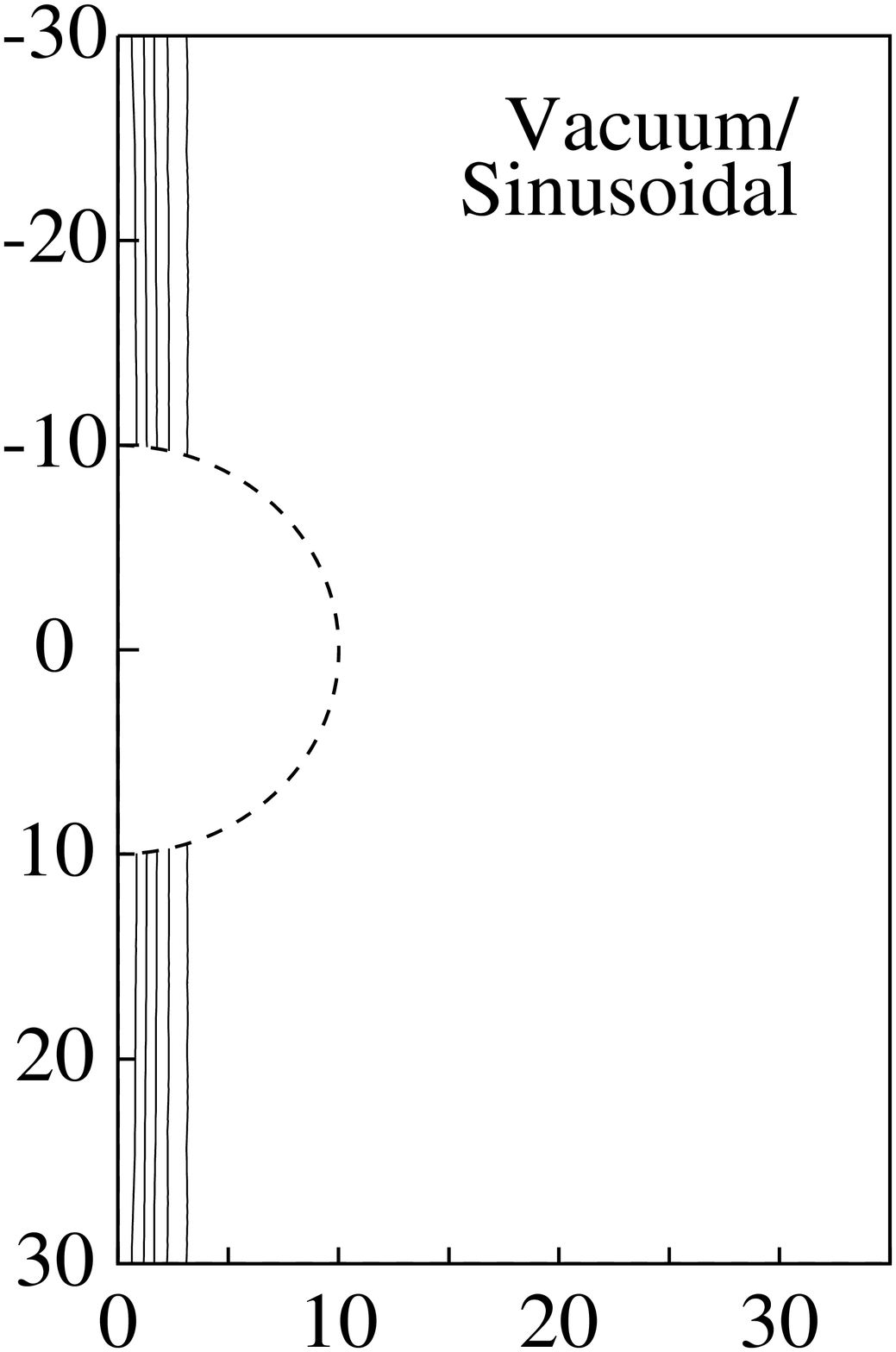,width=3.8cm,angle=180}
  \end{center}
  \caption{A plot of the $P$-contours for $E = 10, N = 1, \beta = 1$ for the
           three initial data sets on the horizon; the sinusoidal and vacuum
	   data sets give the same picture.}
\end{figure}

As can be seen, if we impose the core initial conditions we reproduce 
the results of \cite{CCES}, but if we choose vacuum or sinusoidal
initial conditions for a thin vortex  we get a very different
result and find that the vortex does indeed pierce
the horizon. We have also computed the energies of these solutions to compare
with the plot of \cite{CCES}, and find that the piercing solutions do indeed
have lower energy for large black holes.

Let us examine why we obtain different results to \cite{CCES}.
The answer clearly lies in the initial
conditions chosen, in \cite{CCES} core
initial conditions were used, however, we find that different initial
conditions give different results.
The problem is that for the extremal black hole,
the horizon in the spatial section is singular, albeit a singularity at
infinite proper distance away. Great care must therefore be taken with 
interpreting numerical work, and a variety of different approaches
must be used. We believe the piercing solution is the correct one for
thin strings for the following reasons: First, its energy
is lower, second, it makes more physical sense --
what would the equilibrium solution for the black hole/vortex system be
otherwise? And finally, the numerical data for the piercing solution was
much more robust. Because the horizon is an infinite spacelike distance away, if
$X$ is set to zero there, the method can never lift it from zero, it simply
tries to sweep all the variation in the fields as close to the horizon
as it can to minimise the errors.

\section*{Acknowledgements}
 
Our thanks to Roberto Emparan, Konrad Kuijken and Andrew Sornborger
for useful discussions.

\end{document}